\documentclass{appolb}
\usepackage{epsfig,bm,amsmath,picins}

\def\gs{{\gamma^*}}
\def\qa{{q\bar q}}
\def\G{{\cal G}}

\def\Fs#1{{}\kern-.45em \not \kern-.12em #1\hspace{.2pt}}

\begin{document}
\pagestyle{plain}

\title{Status of the NLO Corrections to the \\Photon Impact Factor%
  \thanks{Presented at the X International Workshop on Deep Inelastic
    Scattering and Related Phenomena (DIS2002), Cracow, 30 April - 4
    May 2002.}}  
\author{Stefan Gieseke\thanks{Supported by the EU
    TMR-Network `QCD and the Deep Structure of Elementary Particles',
    contract number FMRX-CT98-0194 (DG 12-MIHT).}  
  \address{University of Cambridge, Cavendish Laboratory\\
    Madingley Road, Cambridge CB3 0HE, United Kingdom}} 
\maketitle

\begin{abstract}
\noindent
We present the status of the programme of calculating the
next-to-leading order corrections to the virtual photon impact factor.
In particular, we discuss new results for the transversely polarized
photon.  We briefly outline the definition of infrared finite terms
and the subtraction of the leading logarithmic parts.
\end{abstract}


\section{Introduction}
\noindent
The total cross section for the scattering of two highly virtual
photons, having virtualities $Q_1^2$ and $Q_2^2$ at large
centre-of-mass energy $s$ ($s \gg Q_1^2, Q_2^2$) is an excellent
testing ground for the applicability of perturbative QCD in the Regge
limit \cite{BdRLgsgs, brodsky}.  If the energy is high enough to
expect the validity of the Regge limit but not too high in order to
suppress unitarity corrections we expect the $\gs\gs$ cross section to
be described by the BFKL \cite{BFKL} equation.

Besides $\gs\gs$ scattering we are interested in understanding the
$\gs p$ cross section or the structure function $F_2(x, Q^2)$ at small
values of $x = Q^2/s$.  At large values of $Q^2$ we can describe the
contribution from the photon side in perturbation theory, which might
allow us to get insight into the non-perturbative nature of the
proton.  Furthermore, we are able to study the interplay between the
hard and the soft pomeron from the point of view of perturbative QCD.
In addition, we may force a situation that is believed to be purely
perturbative in $\gs p$ scattering as well when we consider the
production of forward jets at HERA \cite{KMSjets, B_etal_fwjets},
similar to the production of Mueller-Navelet jets in $pp$-scattering
\cite{MNjets}.  The coupling of the BFKL ladder to the relevant jet
production vertex at NLO is currently under study \cite{BCV}.

To leading logarithmic accuracy (LLA) the predicted cross section,
based on the BFKL equation, rises too quickly with increasing $s$.
The situation is very different at NLO.  The calculation of NLO
corrections to the BFKL Kernel was initiated in \cite{NLOstart} and
finally completed in \cite{FL, CC}.  The corrections were first seen
to be very large.  However, their size is under control when
additional collinear logarithms are taken into account \cite{CCS,
  salamschool} or when the kinematical conditions are forced to avoid
these extra logarithms (rapidity vetoes) \cite{schmidt, FRSV}.  The
NLO corrections tend to lower the power rise of cross sections to
values that seem to be compatible with the data \cite{BES, expgsgs}.
These studies, however, can at best be viewed as an estimate of higher
order corrections since they do not take care of higher order
corrections to the coupling between external particles, virtual
photons in the $\gs\gs$ case, and the NLO BFKL ladder.

In order to make reliable predictions, being consistent to NLO, the
NLO corrections to the coupling of virtual photons to the exchanged
BFKL ladder, described by the impact factor, has to be taken into
account.  These corrections are currently under study \cite{BGQ, BGK,
  BCGK} and the status of this work is reviewed in this contribution.

\section{The calculational programme}
\noindent
We focus our discussion on the example of $\gs\gs$-scattering which
may serve as the canonical example for a scattering process in
perturbative QCD at very high energy.

\piccaption[]{Regge factorization of the $\gs\gs$ scattering process. 
  \label{fig:reggefact}}
\parpic(5cm,5.5cm)[r]{
  \epsfig{file=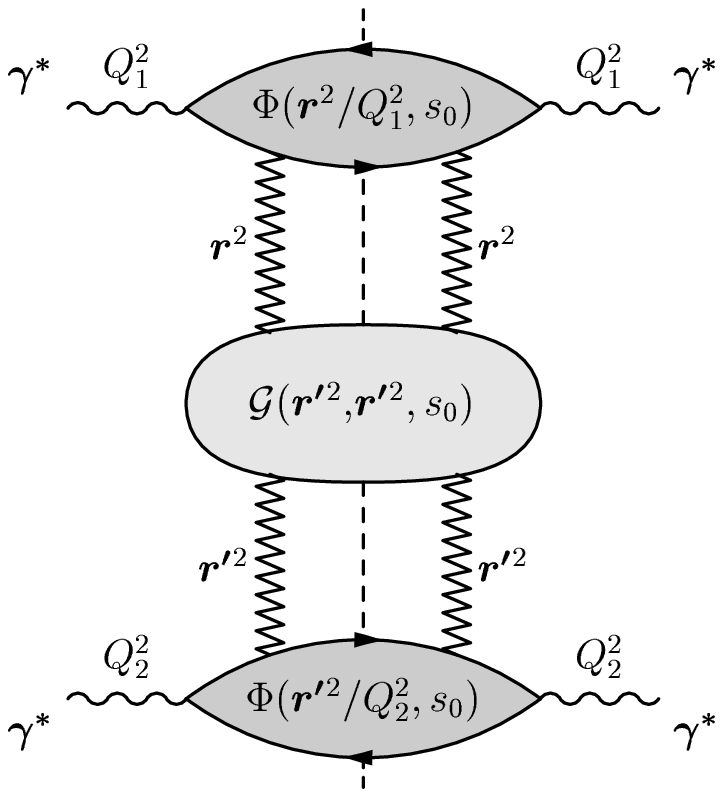,width=5cm}
}
As a result of Regge factorization, always assumed at very high
energy, we can write down the total cross section for
$\gs\gs$-scattering as (cf.\ Fig.~\ref{fig:reggefact})
\begin{equation}
    \label{eq:reggefact}
    \sigma_{\gs\gs}(s) = 
    {\Phi_\gs} \otimes
    {\G_\omega} \otimes
    {\Phi_\gs}\;, 
  \end{equation}
where $\G_\omega (\bm r^2, {\bm r'}^2, s_0)$ is the Green's function for the
exchange of two reggeized gluons, projected into the colour singlet
state, obtained as a solution of the (NLO) BFKL equation.  $\Phi_\gs$
is the impact factor for virtual photons under discussion.  At leading
order, this impact factor (Fig.~\ref{fig:gsif}) is calculated from
cut quark box diagrams: the virtual photon splits into a $\qa$-pair
and the reggeized gluons from the $t$-channel couple to the $\qa$ pair
in all possible ways.

\piccaption[]{The $\gs$ impact factor. \label{fig:gsif}}
\parpic(6.5cm,2.5cm)[r]{
      $\Phi_{\gs} = $\parbox{5cm}{\epsfig{file=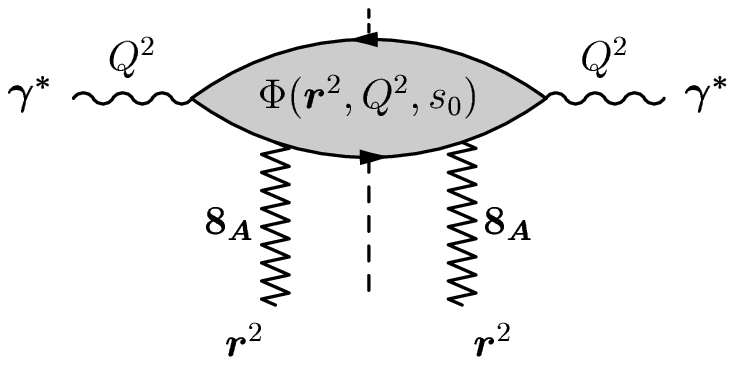,width=5cm}}
}
At NLO we have virtual corrections to the leading order diagrams as
well as contributions with an additional gluon in the intermediate
state (Fig.~\ref{fig:contribif}) and the impact factor at NLO would
look like
\begin{equation}
  \Phi_{\gs}^{(1)} =  
  \int\!\frac{dM_{\qa}^2}{2\pi} d\phi_{\qa}\,
  2\,{\mathrm{Re}}\, 
  \Gamma^{(1)}_{\gs\to\qa}\Gamma^{(0)}_{\gs\to\qa}
  + \int\!\frac{dM_{\qa g}^2}{2\pi} d\phi_{\qa g}
  \left|\Gamma^{(0)}_{\gs\to\qa g}\right|^2\;.
\end{equation}
The terms $\Gamma^{(0)}_{\gs\to\qa}$, $\Gamma^{(1)}_{\gs\to\qa}$ and
$\Gamma^{(0)}_{\gs\to\qa g}$ are the particle-particle-reggeon
vertices that naturally appear at amplitude level as a result of Regge
factorization.  $dM^2_i$ and $d\phi_i$ denote the invariant mass and
the phase space of the respective intermediate states $i=\qa, \qa g$.
\piccaption[]{Contributions to the $\gs$-impact factor at NLO.
  \label{fig:contribif}}
\parpic(8.5cm,2.5cm)[r]{
  \epsfig{file=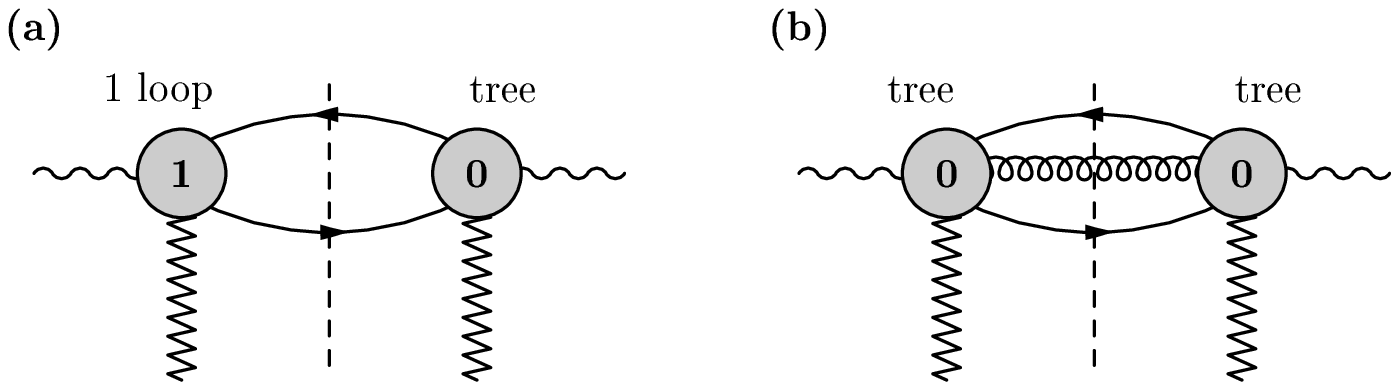,width=8cm}
}

The virtual corrections $\Gamma^{(1)}_{\gs\to\qa}$ have been
calculated in \cite{BGQ}.  We have expressed all loop integrals in
analytic form as an expansion in $\epsilon = (4-D)/2$, quite in
contrast to \cite{FIK} where all integrals are kept as they are.  For
the real corrections, we considered the square of the particle-reggeon
vertex $|\Gamma^{(0)}_{\gs\to\qa g}|^2$, which has been calculated in
\cite{BGK} for longitudinally polarised virtual photons.  In
\cite{BCGK} we complete the real corrections by adding the
contributions from transversely polarised photons.  The real
corrections have been considered in \cite{FIK2} as well, but not in a
very suitable form for the task of finally evaluating the impact
factor.

However, calculating the amplitudes as they are does not quite
complete the task.  The individual contributions are still infrared
divergent and have to be combined in order to get the expected
cancellation that has been shown previously \cite{FM}.  At the same
time, one has to consider the subtraction of leading logarithmic
terms.  These are present in the virtual corrections and proportional
to the well-known LLA gluon trajectory function. In the real
contribution to the impact factor they arise as the additional gluon
is emitted with a large rapidity separation to the $\qa$-pair.  Both
of these LLA-terms are individually infrared divergent as well as the
emitted gluon becomes soft.  In \cite{BCGK} we have extracted the
infrared divergent contributions from real and virtual corrections and
defined suitable subtraction terms.  The difference of our result and
the respective subtraction term is finite upon integration over the
gluon phase space.  Re-adding the subtracted terms with the
integrations over the gluon phase space performed, explicitly allows
us to exhibit the infrared divergences and cancel them successfully
against those from the virtual corrections.  The subtraction of the
leading logarithmic terms induces a scale $s_0$ which can be
translated into a rapidity cutoff beyond which the emitted gluon will
belong to the leading logarithmic term.  However, since the particular
choice of this scale is arbitrary, the NLL impact factors will depend
on it.  This dependence was irrelevant at LLA, since a change in the
scale
\begin{equation}
  \label{eq:log}
  \ln\frac{s}{s_0} = \ln\frac{s}{s_1} + \ln\frac{s_1}{s_0} = 
  \ln\frac{s}{s_1} + \mathrm{NLLA}
\end{equation}
is of higher order w.r.t. the LLA.  These NLLA terms are now taken
care of and may be phenomenologically important.

\section{Outlook and Conclusions}
\noindent
Besides the above discussion of the NLO impact factor, our
calculations have the potential to give further insight into the
photon wave function picture. This picture, in conjunction with the
saturation model has been applied successfully to the description of
both deep-inelastic and diffractive scattering cross sections at HERA,
e.g.\ \cite{GBW1, GBW2}.  First steps in this direction have been done
in \cite{BGK}, showing that an extension of the current picture to a
higher $\qa g$ Fock-state of the virtual photon is in principle
possible.  Further steps in this direction include a consistent
treatment of infrared divergences in configuration space and remain
to be done.

In order to complete the calculation of the impact factor we have to
calculate the phase space integrals over the remaining infrared finite
terms, defined in \cite{BCGK}.  We will express the phase space
integrals in terms of a set of standard integrals.  For first
phenomenological applications this might best be done numerically.

With these results we will be able to calculate the $\gs\gs$ cross
section to NLL accuracy.  In combination with the NLO jet vertex
\cite{BCV} there will be an interesting variety of phenomenological
applications for the NLO BFKL equation.

\end{document}